\documentclass[reprint,
amsmath,amssymb,
aps,
]{revtex4-2}
\usepackage{graphicx}
\usepackage{dcolumn}
\usepackage{bm}
\usepackage{here}

\begin{document}

\title{
Persistence of fermionic spin excitations\\
through a genuine Mott transition in $\kappa$-type organics
}

\author{
S. Imajo$^{1,2,*}$,$\thanks{imajo@issp.u-tokyo.ac.jp}$
N. Kato$^{1}$,
R. J. Marckwardt$^{1}$,
E. Yesil$^{1}$,
H. Akutsu$^{1}$,
and
Y. Nakazawa$^{1}$
}
\affiliation{
$^1$Graduate School of Science, Osaka University, Toyonaka, Osaka 560-0043, Japan\\
$^2$Institute for Solid State Physics, University of Tokyo, Kashiwa, Chiba 277-8581, Japan
}

\date{\today}

\begin{abstract}
We investigate the continuous variation of electronic states from a Fermi liquid to a quantum spin liquid induced by chemical substitution in the organic dimer-Mott system of $\kappa$-[(BEDSe-TTF)$_x$(BEDT-TTF)$_{1-x}$]$_2$Cu[N(CN)$_2$]Br.
Electrical transport measurements reveal that the mixing of BEDSe-TTF into the BEDT-TTF layers induces a quantum Mott transition around $x$=0.10.
Although a charge gap disappears at this point, magnetic susceptibility and heat capacity measurements indicate that fermionic low-energy excitations remain in the insulating salts, suggesting that fermionic spin excitations persist.
We propose that the transition can be classified into a genuine Mott transition.
\end{abstract}

\maketitle
\begin{figure}[t]
\begin{center}
\includegraphics[width=0.95\hsize,clip]{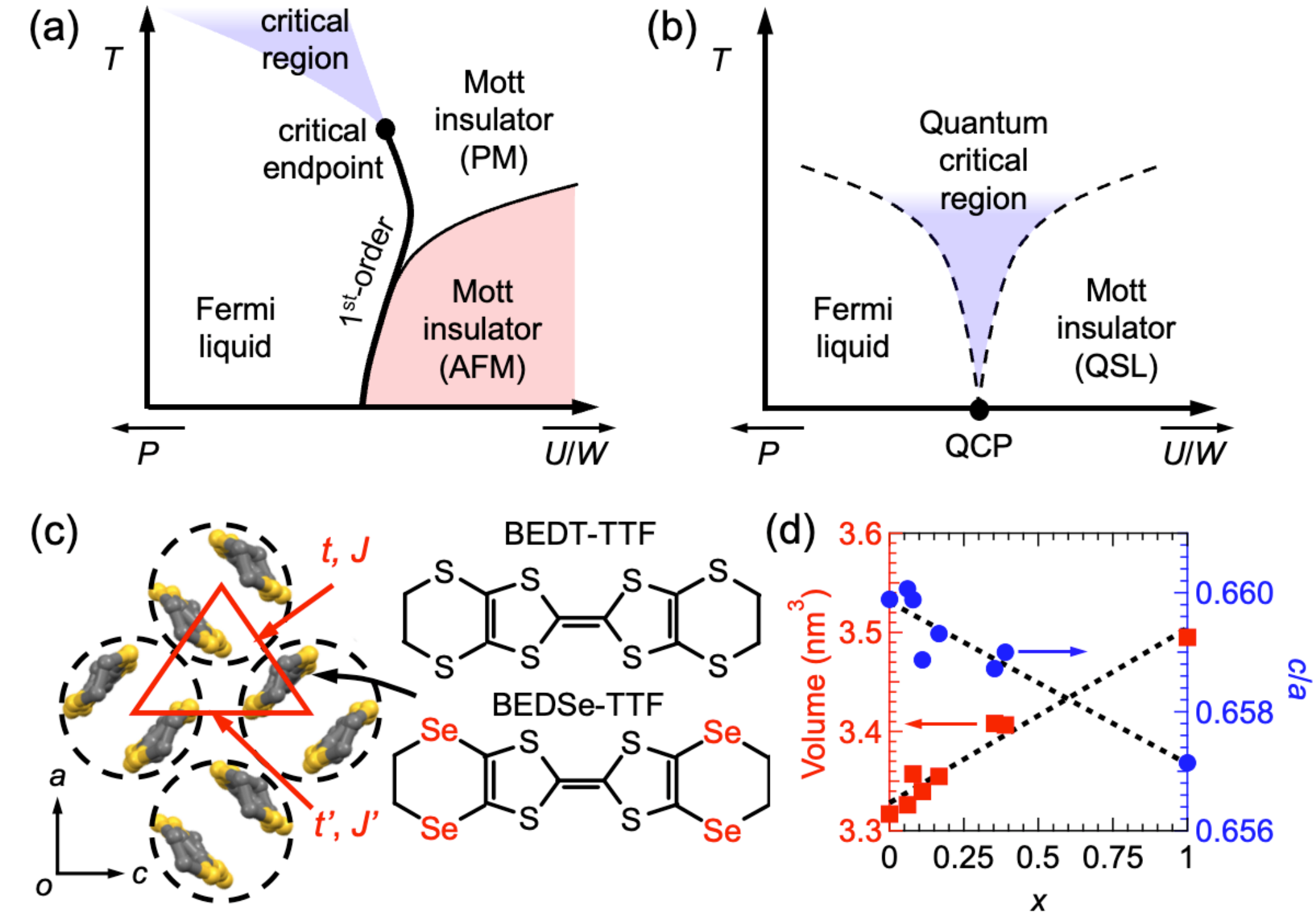}
\end{center}
\caption{
(a),(b) Schematic phase diagrams near pressure-induced Mott transition, accompanied by (a) antiferromagnetic transition and (b) no magnetic transition.
(c) Arrangement of the donor molecules in $\kappa$-type organics.
Molecular dimers form distorted triangular lattice, characterized by the transfer integral ratio $t^{\prime}$/$t$.
Right figure shows chemical structural formulae of BEDT-TTF and BEDSe-TTF.
(d) Mixing ratio $x$ dependence of unit-cell volume $V$ and the ratio of in-plane unit-cell length $c$/$a$ of $\kappa$-[(BEDSe-TTF)$_x$(BEDT-TTF)$_{1-x}$]$_2$Cu[N(CN)$_2$]Br at room temperature.
Data point for $x$=1 is obtained from Ref.~\cite{10}.
Dotted lines are linear fits to these parameters.
}
\label{fig1}
\end{figure}

  Electron correlations can trigger a metal-insulator transition, known as a Mott transition, when Coulomb repulsion $U$ exceeds bandwidth $W$ in a half-filled electronic state\cite{1}.
Because the magnitude of the ratio $U$/$W$ determines whether an electron is localized on each site or not, the transition is primarily related to the charge degrees of freedom, namely the presence or absence of the itinerancy of electrons.
However, in most Mott systems with half-filling states, such as V$_2$O$_3$\cite{2}, high-$T_{\rm c}$ cuprate\cite{3}, and typical dimer-Mott organics\cite{4}, the Mott transition is simultaneously accompanied by antiferromagnetic ordering of spins on the localized electrons.
In this case, the Mott transition discontinuously divides the metallic phase from the insulating phase, as shown in Fig.~\ref{fig1}a.
To discuss the Mott transition genuinely, the Mott transition should not involve a change in the magnetic degrees of freedom.
Therefore, the Mott insulating state must also be a magnetically non-ordered state, such as a quantum spin liquid (QSL), in which spins in strong electron correlations fluctuate even at low temperatures.

  In a genuine Mott transition from a conventional Fermi liquid (FL) to a QSL, excitations of the spin sector of electrons, namely charge-neutral spinons, are expected to remain and may possibly be delocalized even in the QSL Mott insulator.
It has been theoretically predicted that that the QSL Mott transition exhibits some peculiar features, such as a continuous variation in physical quantities (Fig.~\ref{fig1}b) and the presence of a spinon Fermi surface in the Mott state\cite{5}.
A recent study\cite{6} detected a possible indication of the quasi-continuous Mott transition based on electrical transport measurements under gas-pressure control.
A spectroscopic study on various organic QSLs reveals that the geneuine Mott transition is understood in a generic phas diagram using $U$/$W$ and $T$/$W$\cite{6p3}.
Although some previous studies\cite{7,8,9} suggest that organic QSL materials show gapless excitations as if the putative spinon Fermi surface exists, a detailed process pertaining to the emergence of the gapless spin excitations through the genuine Mott transition has yet to be reported.
Moreover, the ground state and the presence of the gapless excitations are still controversial in the prime QSL candidate $\kappa$-(BEDT-TTF)$_2$Cu$_2$(CN)$_3$, in which contradictions have been not resolved yet\cite{7,16,9p3,9p6}.
Therefore, investigations of new $\kappa$-type QSL systems may be useful to discuss the QSL state and the genuine Mott transition.
Herein, we first report a continuous variation in the physical properties of $\kappa$-[(BEDSe-TTF)$_x$(BEDT-TTF)$_{1-x}$]$_2$Cu[N(CN)$_2$]Br (BEDT-TTF and BEDSe-TTF denote bis(ethlenedithio-tetrathiafulvalene) and bis(ethlenediselena-tetrathiafulvalene), respectively, as illustrated in Fig.~\ref{fig1}c).

\begin{table*}
  \caption{Crystallographic data.}
  \label{tab1}
  \begin{center}
  \begin{tabular}{c||cccccc}
    \hline \hline
    ~ & $x$=0.06  & $x$=0.08  &  $x$=0.11  &  $x$=0.17  &  $x$=0.35  &  $x$=0.39\\
    \hline \hline
    $x$ ($\%$)  & 6.04  & 7.85  & 11.0  & 16.6  & 35.3  & 38.9\\
    Formula  & C$_{22}$H$_{16}$N$_{3}$S$_{15.52}$  & C$_{22}$H$_{16}$N$_{3}$S$_{15.38}$  & C$_{22}$H$_{16}$N$_{3}$S$_{15.12}$  & C$_{22}$H$_{16}$N$_{3}$S$_{14.67}$ & C$_{22}$H$_{16}$N$_{3}$S$_{13.17}$  & C$_{22}$H$_{16}$N$_{3}$S$_{12.89}$\\
    ~  & Se$_{0.48}$CuBr  & Se$_{0.62}$CuBr  & Se$_{0.88}$CuBr  & Se$_{1.33}$CuBr  & Se$_{2.83}$CuBr  & Se$_{3.11}$CuBr\\
    Fw  & 1001.31  & 1007.88  & 1020.07  & 1041.18  & 1111.53  & 1124.66\\
    Space group  & $Pnma$  & $Pnma$  & $Pnma$  & $Pnma$  & $Pnma$  & $Pnma$\\
    $a$ (${\rm \AA}$)  & 12.9576(7)  & 12.9958(8)  & 12.9884(3)  & 13.0055(4)  & 13.0836(5)  & 13.0766(3)\\
    $b$ (${\rm \AA}$)  & 30.0142(13)  & 30.125(2)  & 30.0457(7)  & 30.0845(10)  & 30.226(2)  & 30.2305(6)\\
    $c$ (${\rm \AA}$)  & 8.5529(4)  & 8.5757(5)  & 8.5577(2)  & 8.5747(2)  & 8.6185(3)  & 8.61749(17)\\
    $V$ (${\rm \AA}$$^3$)  & 3326.3(3)  & 3357.4(4)  & 3339.59(14)  & 3354.97(17)  & 3408.3(3)  & 3406.59(12)\\
    $Z$  & 4  & 4  & 4  & 4  & 4  & 4\\
    $T$ (K)  & 290  & 296  & 293  & 296  & 290  & 290\\
    $d$$_{\rm calc}$ (g cm$^{-1}$)  & 1.999  & 1.994  & 2.029  & 2.061  & 2.166  & 2.193\\
    $\mu$ (mm$^{-1}$)  & 3.386  & 3.498  & 3.784  & 4.227  & 5.671  & 5.956\\
    $F$(000)  & 1990.56  & 2000.64  & 2019.36  & 2051.76  & 2159.76  & 2179.92\\
    2$\theta$range($^{\circ}$)  & 4-55  & 4-55  & 4-55  & 4-55  & 4-55  & 4-55\\
    Total ref.  & 30187  & 29359  & 30210  & 29691  & 23411  & 30895\\
    Unique ref.  & 3881  & 3906  & 3886  & 3896  & 3889  & 3957\\
    $R$$_{\rm int}$  & 0.1397  & 0.1753  & 0.1332  & 0.0560  & 0.0793  & 0.0759\\
    Parameters  & 218  & 218  & 218  & 218  & 218  & 218\\
    $R$$_{1}$ ($I$$>$2$\sigma$($I$))  & 0.069  & 0.109  & 0.051  & 0.040  & 0.0609  & 0.051\\
    w$R$$_{2}$ (all data)  & 0.204  & 0.125  & 0.135  & 0.099  & 0.143  & 0.146\\
    $S$  & 1.041  & 1.076  & 1.059  & 1.022  & 1.038  & 1.036\\
    $\Delta$$\rho$$_{\rm max}$ (e ${\rm \AA}$$^3$)  & 1.68  & 2.15  & 1.08  & 0.78  & 1.33  & 1.26\\
    $\Delta$$\rho$$_{\rm min}$ (e ${\rm \AA}$$^3$)  & -0.57  & -0.78  & -0.86  & -0.57  & -0.78  & -0.78\\
  \end{tabular}
  \end{center}
\end{table*}
The pristine salt $\kappa$-(BEDT-TTF)$_2$Cu[N(CN)$_2$]Br (abbreviated BEDT-Br) is a well-known organic superconductor with a superconducting transition temperature $T_{\rm c}$$\sim$12~K\cite{4}.
As shown in Fig.~\ref{fig1}c, BEDSe-TTF is a molecule in which the four S atoms in the two outer six-membered rings of BEDT-TTF are replaced by Se.
This replacement results in an elongation of the intermolecular distance without an increase in the overlap integrals in the crystals because of the lower $\pi$-electron density on the outer chalcogen sites.
$U$ of the dimers, almost proportional to the transfer integral of the dimerization in this system, is hardly changed by this substitution\cite{10}, and therefore, this substitution works as a negative chemical pressure for the electronic state.
In fact, $\kappa$-(BEDSe-TTF)$_2$Cu[N(CN)$_2$]Br (BEST-Br) shows insulating behavior due to the large $U$/$W$ at ambient pressure and superconducting behavior, similar to BEDT-Br, at high pressure\cite{10}.
Interestingly, the insulating state does not exhibit a distinct magnetic transition down to 2~K, in marked contrast to $\kappa$-(BEDT-TTF)$_2$Cu[N(CN)$_2$]Cl\cite{11} and $\kappa$-(d8-BEDT-TTF)$_2$Cu[N(CN)$_2$]Br\cite{12}, which are also the analogous salts with antiferromagnetic (AFM) transitions at $T_{\rm N}$$\sim$23~K and $\sim$15~K.
The absence of an AFM transition and the temperature dependence of the magnetic susceptibility in BEST-Br imply that its ground state is the QSL state.
Therefore, tuning the ratio $x$ of $\kappa$-[(BEDSe-TTF)$_x$(BEDT-TTF)$_{1-x}$]$_2$Cu[N(CN)$_2$]Br may enable access to the QSL Mott transition.
Previous works regarding $\kappa$-[(BEDSe-TTF)$_x$(BEDT-TTF)$_{1-x}$]$_2$Cu[N(CN)$_2$]Br only reported the suppression of $T_{\rm c}$ with increasing $x$\cite{13,14,15}, whereas details concerning the ground state were not provided.
In this study, we investigated $\kappa$-[(BEDSe-TTF)$_x$(BEDT-TTF)$_{1-x}$]$_2$Cu[N(CN)$_2$]Br to explore for the Mott transition and understand its nature.

Single crystals of $\kappa$-[(BEDSe-TTF)$_x$(BEDT-TTF)$_{1-x}$]$_2$Cu[N(CN)$_2$]Br were prepared by electrochemical oxidations of mixed donor molecules BEDT-TTF and BEDSe-TTF.
Note that our synthetic method did not yield salts with $x$$>$0.5.
From X-ray crystal structure analyses, the mixing ratio $x$ of the samples was determined by the average occupancy probability of the Se atom in the outer chalcogen sites.
No peak splitting of X-ray diffraction confirmed that the mixed crystals are macroscopically homogeneous as the lattice parameters almost linearly change depending on $x$ (Fig.~\ref{fig1}d).
The detailed structural parameters of the measured crystals are shown in Table~\ref{tab1}.
For the resistivity measurements, we employed the typical four-probe method and measured the in-plane resistance.
The magnetic susceptibility measurements were performed by a SQUID magnetometer (MPMS2, Quantum Design) at 1~T with polycrystalline samples weighing about $\sim$5~mg.
The analyses of the obtained data are described in Supplemental Materials\cite{Suppl}.
Using a homemade high-resolution calorimeter\cite{S1}, the low-temperature heat capacity measurements of the single crystals, whose mass is typically about 300~$\mu$g, were carried out in a 9~T superconducting magnet with a $^3$He cryostat.

Before discussing the present results, we must first need to discuss the absence of magnetic transitions reported in the pure salt BEST-Br\cite{10} ($x$=1), which implies the QSL ground state.
The QSL state is typically observed where the systems possess geometrical frustration of spins\cite{31p3} and/or randomness interrupting magnetic interactions\cite{31p6}.
For the effect of the randomness, the similarities of the structures and molecules in forming the crystals between BEST-Br and the AFM salts, $\kappa$-(BEDT-TTF)$_2$Cu[N(CN)$_2$]Cl\cite{11} and $\kappa$-(d8-BEDT-TTF)$_2$Cu[N(CN)$_2$]Br\cite{12}, indicate that the magnitude of randomness in BEST-Br should be comparable to that of the AFM salts.
Therefore, the frustration of the dimer lattice should also be considered as a factor that suppresses the AFM in BEST-Br.
Based on the lattice parameter ratio in the conducting plane $c$/$a$ shown in Fig.~\ref{fig1}d,the smaller $c$/$a$ for BEST-Br indicates that the dimer lattice of BEST-Br have a larger ratio of the transfer integrals $t^{\prime}$/$t$ (Fig.~\ref{fig1}c) than that of BEDT-Br.
In fact, using the H$\ddot{\rm u}$ckel tight-binding band calculation method, the $t^{\prime}$/$t$ of BEST-Br is 1.25\cite{10}, which is higher than that of BEDT-Br, 0.68\cite{10,32}, and that of the QSL salt $\kappa$-(BEDT-TTF)$_2$Cu$_2$(CN)$_3$, 1.06\cite{32}.
Note that the value of $t^{\prime}$/$t$ depends on the calculation method\cite{32,33,33p4}.
Although $t^{\prime}$/$t$ obtained by the more precise first-principles DFT calculation\cite{33,33p4} tends to smaller than that given by the H$\ddot{\rm u}$ckel calculation in Ref.~\cite{10,32}, the relative comparison of $t^{\prime}$/$t$ in $\kappa$-(BEDT-TTF)$_2$X salts are often valid even in the framework of the H$\ddot{\rm u}$ckel calculation.
In Ref.~\cite{33p4}, they show $t^{\prime}$/$t$ for $\kappa$-(BEDT-TTF)$_2$Cu$_2$(CN)$_3$, $\kappa$-(BEDT-TTF)$_2$Cu(NCS)$_2$, $\kappa$-(BEDT-TTF)$_2$Cu[N(CN)$_2$]Cl, and $\kappa$-(d8-BEDT-TTF)$_2$Cu[N(CN)$_2$]Br, which are 0.8-0.9, 0.52, 0.52, and 0.45, in the DFT calculations, and 1.06, 0.80, 0.75, and 0.69 in the H$\ddot{\rm u}$ckel calculations, respectively.
Considering the overestimation of 0.2-0.3 in the H$\ddot{\rm u}$ckel calculation and the relative comparison, BEST-Br possibly have a relatively strong geometric frustration with $t^{\prime}$/$t$$\sim$1.

\begin{figure}[h]
\begin{center}
\includegraphics[width=\hsize]{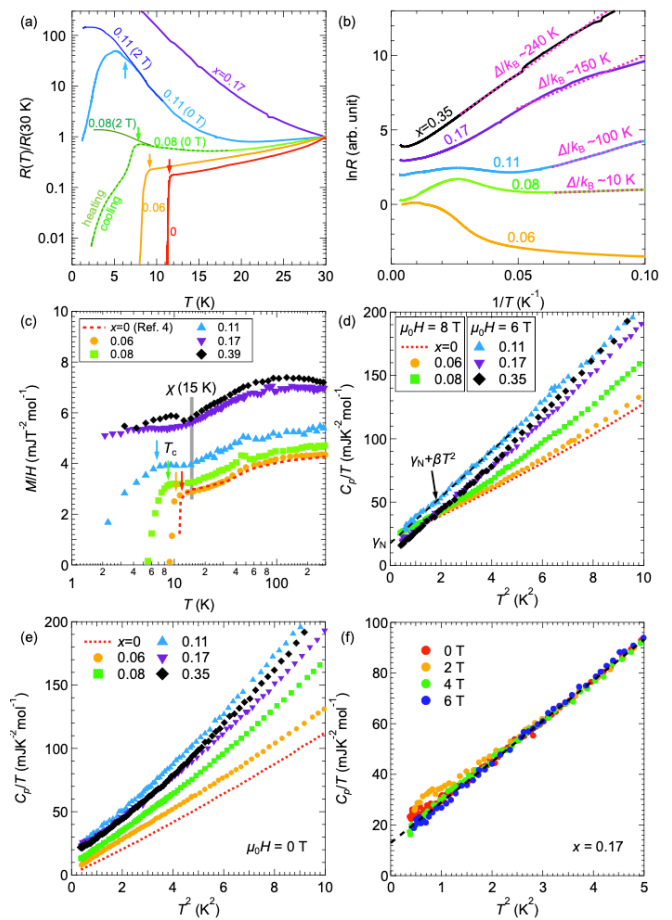}
\end{center}
\caption{
(a) Temperature dependence of reduced resistivity $R$($T$)/$R$(30~K) of mixed salts $\kappa$-[(BEDSe-TTF)$_x$(BEDT-TTF)$_{1-x}$]$_2$Cu[N(CN)$_2$]Br ($x$=0, 0.06, 0.08, 0.11, and 0.17).
For $x$=0.06 and 0.08, data at 2~T are also plotted.
Arrows denote superconducting transition temperature $T_{\rm c}$.
Dotted curve on the $x$=0.08 indicates that data do not show hysteresis depending on heating and cooling.
(b) Arrhenius plot of the resistance.
Dotted lines are linear fits using the respective charge gap $\Delta$/$k_{\rm B}$.
(c) Magnetic susceptibility $M$/$H$ at 1~T as a function of temperature.
Data for $x$=0 is obtained from Ref.~\cite{4}.
(d) $C_p$/$T$ vs. $T^2$ for each salt in a magnetic field.
Dashed line shows a fitting to $C_p$/$T$=$\gamma$$_{\rm N}$+$\beta$$T^2$, where $\gamma$$_{\rm N}$ and $\beta$ are the electronic and lattice heat capacity coefficients, respectively.
(e) Zero-field heat capacity for each mixed salt plotted as $C_p$/$T$ vs $T^2$.
We additionally show the data for $x$=0 taken from the literature\cite{S2}.
(f) Low-temperature heat capacity of the salt $x$=0.17 at 0, 2, 4, and 6~T.
}
\label{fig2}
\end{figure}

  Figure~\ref{fig2}a shows the temperature dependence of the reduced resistivity $R$($T$)/$R$(30~K) of the crystals with different stoichiometries ($x$=0, 0.06, 0.08, 0.11, and 0.17).
As indicated by the arrows, the salts with $x$$\leq$0.11 exhibit a superconducting transition and $T_{\rm c}$ decreases as $x$ increase, which is consistent with the results of the earlier studies\cite{13,14}.
The imperfect resistivity drop for the $x$=0.08 and 0.11 salts indicates the percolation of the superconductivity, probably induced by inhomogeneity.
Above $T_{\rm c}$, the compounds with $x$=0 and 0.06 show the metallic behavior, whereas the compound with $x$=0.08 exhibits an almost temperature-independent behavior.
The compound with $x$=0.11 indicates a weak insulating behavior and then undergoes a superconducting transition at $\sim$5~K.
The broad and incomplete resistivity decrease for the salt with $x$=0.11 indicates that it is primarily insulating and has slight percolative superconductivity, which is due to the local inhomogeneity of $x$.
The salts with $x$$\geq$0.17 exhibit only an insulating nature.
This change indicates that the salts with $x$=0.08-0.11 are located at the verge of the phase boundary of the metal-insulator transition.
The Arrhenius plot of the resistivity is shown in Fig.~\ref{fig2}b.
As shown, the salts with $x$=0.08-0.11 are near the metal-insulator critical point because of the inconspicuous charge gap $\Delta$/$k_{\rm B}$$\sim$10$^{\rm 1}$-10$^{\rm 2}$~K, obtained by linear fitting to 2d(ln$R$)/d(1/$T$).
In the insulators, a continuous development of the charge gap $\Delta$/$k_{\rm B}$ with increasing $x$ is observed.

The magnetic susceptibility $\chi$=$M$/$H$ obtained at 1~T is shown in Fig.~\ref{fig2}c as a function of temperature.
Susceptibility is deduced by subtracting the contribution of the impurity spins behaving as a Curie-type paramagnetic component\cite{Suppl}.
The superconducting transition is observed in the $x$$\leq$0.11 salts , and its temperature is reduced by the substitution, consistent with the resistivity results.
For the insulating salts with $x$$\geq$0.11, $T$-independent susceptibility is clearly observed for $T$$\rightarrow$0~K ($\chi$$_0$).
In addition, magnetic ordering is absent in the Mott insulating region above the concentration of the quantum Mott transition (QMT).
The existence of $\chi$$_0$ implies that the gapless spin excitations are present even in the Mott insulator, as in the case of other organic QSL materials\cite{7,8,9}.
The salts on the metal side naturally exhibit a finite $\chi$$_0$ (estimated from the value above $T_{\rm c}$), attributed to the typical Pauli paramagnetic contribution proportional to the density of states at the Fermi level $D$($E_{\rm F}$).

Figure~\ref{fig2}d shows the heat capacity data of the normal state obtained in magnetic fields, plotted as $C_p$/$T$ vs. $T^2$.
The applied magnetic fields are sufficient to suppress the superconductivity observed in the $x$$\leq$0.11 salts.
Using the formula for the low-temperature heat capacity of normal FLs, $C_p$/$T$=$\gamma$$_{\rm N}$+$\beta$$T^2$, the intercept and the slope of this plot correspond to the electronic heat capacity coefficient $\gamma$$_{\rm N}$ and lattice heat capacity coefficient $\beta$, respectively.
It is clear that all of the salts have finite $\gamma$$_{\rm N}$ values, including in the Mott insulating salts.
Similar to the case of the other $\kappa$-type organic QSLs\cite{16,17,18}, this finite $\gamma$$_{\rm N}$ also indicates that the Mott insulating state contains gapless excitations, which is consistent with the finite $\chi$$_0$ in the susceptibility.
Additionally, in Fig.~\ref{fig2}e, the zero-field heat capacity is shown together with the data for $x$=0\cite{S2}.
At zero field, the electronic heat capacity of the salts showing the bulk superconductivity decreases due to the formation of the superconducting energy gap.
The pristine salt $x$=0 shows almost zero electronic heat capacity coefficient at zero field $\gamma$$^{\ast}$ whereas the mixed salts showing the superconductivity have finite values of $\gamma$$^{\ast}$.
As discussed in Ref.~\cite{15}, this is more likely due to the impurity scattering at superconducting gap nodes because the present superconductivity is classified into $d$-wave symmetry with line nodes\cite{S4}.
For the higher $x$ salts, a small amount of extrinsic contribution arising from some impurities are included, as displayed in Fig.~\ref{fig2}f.
Since this component behaves as a Schottky anomaly and is smeared out at higher fields ($>$ 5~T), $\gamma$$_{\rm N}$ and $\beta$ are estimated with the data in the field-independent region in Fig.~\ref{fig2}d.

\begin{figure}
\begin{center}
\includegraphics[width=\hsize,clip]{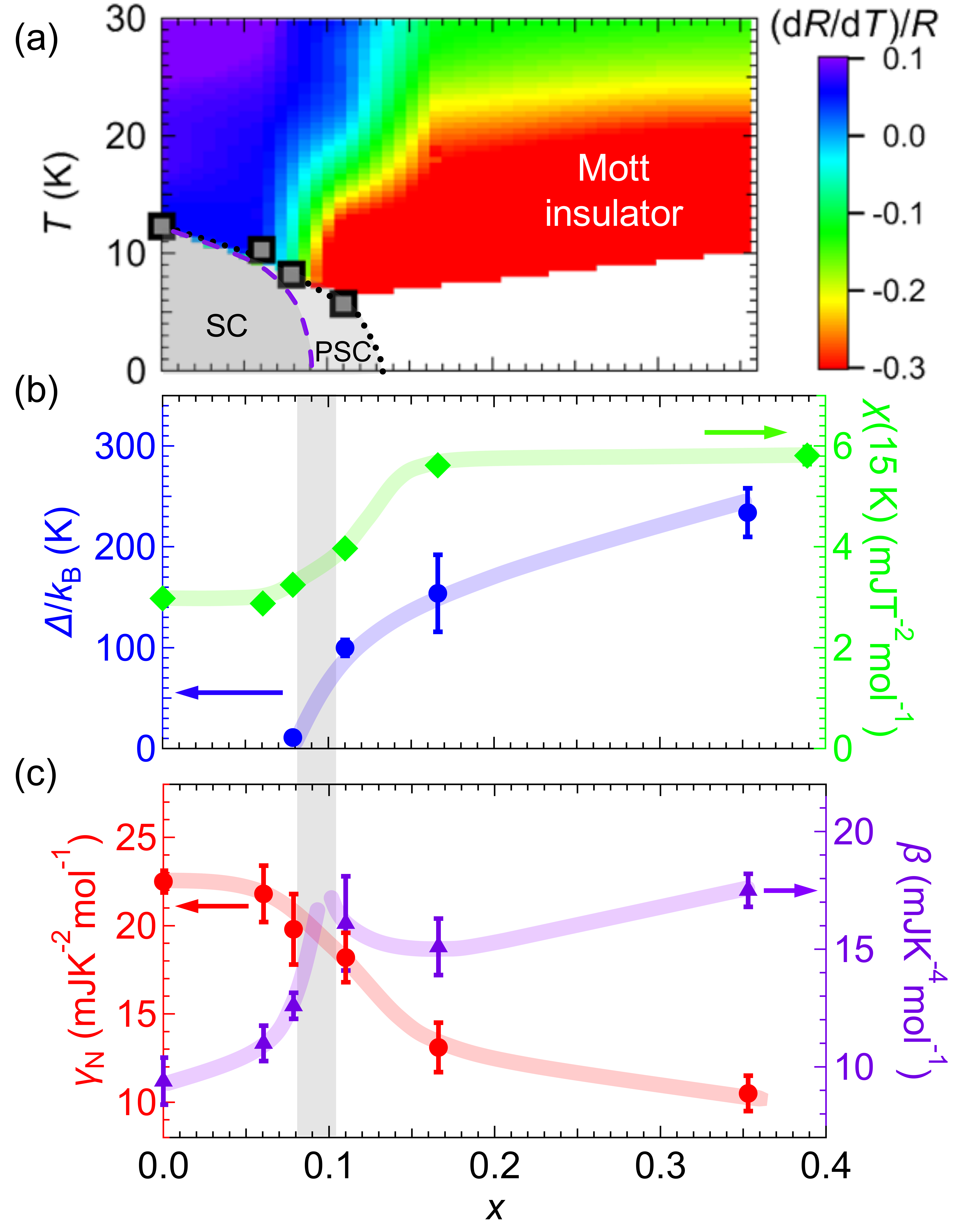}
\end{center}
\caption{
Substitution ratio $x$ dependence of physical parameters: (a) (d$R$/d$T$)/$R$, (b) $\Delta$/$k_{\rm B}$ and $\chi$(15~K), (c) $\gamma$$_{\rm N}$ and $\beta$.
To clarify superconducting and PSC regions, the dotted curve with $T_{\rm c}$ (black box) and dashed curve are also plotted in (a).
Light blue region in (a) roughly corresponds to metal-insulator crossover area (d$R$/d$T$)/$R$=0.
Gray shaded areas in (b) and (c) represent possible position of QMT.
Translucent curves superimposed over data points provide visual guide.
}
\label{fig3}
\end{figure}

  To determine the variation in the physical properties of the ground states, we plot the $x$ dependences of the physical parameters, as shown in Fig.~\ref{fig3}.
First, to characterize the change in the metallicity, the rate of change of resistivity $R$ with temperature (d$R$/d$T$)/$R$ is shown in Fig.~\ref{fig3}a as a contour plot.
In addition, we plot $T_{\rm c}$ (the squares) to denote the superconducting region.
This contour plot shows that the metallicity fades continuously into the insulator side.
The metal-insulator crossover line, roughly indicated by the light blue region, is known as the quantum Widom line\cite{6p3,40p5}, and its extrapolation suggests that the QMT should exist around $x$=0.10, although the percolated superconductivity (PSC) exists and conceals the exact position of the QMT.
Since the disorder effect by donor mixing of $x$$\sim$0.1 reduces $T_{\rm c}$ of the superconductivity in this system to about 90$\%$\cite{38,40p7}, the main reason for the fading out of the superconductivity is considered to be due to the proximity to the QMT.
The percolation simply originates from the local inhomogeneity of $x$ because of the absence of discontinuous resistivity jumps and hysteresis.
If the electronic inhomogeneity is induced by the phase separation appearing near the typical Mott transition, these first-order characteristics must be observed as in the case of the other $\kappa$-type salts\cite{41,18p5}.
Considering some distribution of $x$ and the superconducting volume fraction\cite{Suppl}, it is expected that the bulk superconductivity disappears around $x$=0.1 and the PSC slightly survives nearby, as shown in Fig.~\ref{fig3}a.
In Fig.~\ref{fig3}b, $\Delta$/$k_{\rm B}$ (blue, left axis) is shown.
The finite $\Delta$/$k_{\rm B}$ emerges from the QMT and increases with $x$.
The absence of the discontinuity and the gradual opening of the charge gap imply that the Mott transition is not a first-order one but a second-order one.

  To further investigate the characteristics of this possible continuous transition from the viewpoint of low energy excitations, the obtained $\chi$(15~K) is additionally plotted in Fig.~\ref{fig3}b.
Note that we here evaluate $\chi$(15~K) instead of $\chi$$_0$ in the zero-temperature limit because the superconductivity in the lower-$x$ salts conceals the low-temperature excitations of the normal state.
As shown in Fig.~\ref{fig2}c, since the change in $\chi$ is not large below 15~K, the low-energy magnetic excitations can be roughly discussed using $\chi$(15~K).
In the case of the organic materials, the constant $\chi$$_0$ originates from the Pauli paramagnetic susceptibility, which is one of the features of an FL, because the van Vleck contribution of the light atoms in organics is negligible.
The finite $\chi$ observed at the lowest temperatures for $x$=0.17 and 0.39 is an exact evidence for the presence of fermionic spin excitations, namely spinons, in the insulating salts.
The step-like behavior around $x$=0.1 implies that $\chi$$_0$ is affected by the Mott transition.
The reported 1/$T_{1}$$T$ of $\kappa$-(BEDT-TTF)$_2$Cu$_2$(CN)$_3$ also exhibits a similar behavior near the QSL Mott transition\cite{20}.
This behavior may be attributable to the formation of spinon quasiparticles from electrons when the spin-charge separation occurs.
In the metal phase, the itinerant electrons form an energy band, characterized by the hopping amplitude $t_{\rm electron}$ (=$W$/4).
Meanwhile, when assuming that the spinons in the insulator phase organize an energy band with the Fermi surface, the hopping of the spinons $t_{\rm spinon}$ should be expressed by the Heisenberg exchange energy $J$ as $t_{\rm spinon}$$\sim$$J$ (=$t_{\rm electron}$$^2$/$U$) when the spinons are delocalized\cite{21}.
The difference in the band widths of the spinons and electrons may cause the variation in $\chi$ around the QMT where the origin of the fermionic excitations switches.

  To discuss the excitations arising from the charge sector as well as the spin sector from a thermodynamic viewpoint, we present $\gamma$$_{\rm N}$ in Fig.~\ref{fig3}c because $\gamma$$_{\rm N}$ contains all the low-energy excitations of electrons.
The difference between $\gamma$$_{\rm N}$ and $\chi$$_0$ is attributable to the contribution from the charge sector.
For the Mott transition accompanied by an AFM transition, $\gamma$$_{\rm N}$ decreases significantly near the Mott transition and reaches zero inside the Mott state\cite{4,22} because the fermionic spin excitations are simultaneously suppressed by the formation of long-range magnetic ordering.
By contrast, the present Mott transition results in a moderate and small step-like decrease in $\gamma$$_{\rm N}$ to a finite value, which is due to the disappearance of only the charge excitations without the loss of the fermionic spin excitations; this provides convincing evidence for the genuine Mott transition.
The diverging $\gamma$$_{\rm N}$ around the QMT, which is typical behavior for quantum criticality, is absent; however, the result is consistent with a prediction based on the phenomenological Landau-like low-energy theory\cite{23}, in that $\gamma$$_{\rm N}$ may not diverge at the QSL Mott transition.
Nevertheless, $\beta$ (Fig.~\ref{fig3}c) seems to show the possible diverging behavior.
Considering the error bars, it is difficult to discuss the discontinuity and diverging trend precisely, however, it is confirmed the larger $\beta$, namely the lattice softening, above $x$=0.1.
Around the Mott critical endpoint, a substantial lattice softening, known as the critical elasticity, is observed\cite{24,24p5}.
Since $\beta$ is the parameter characterizing the acoustic phonons in a low-energy limit, this result also indicates that the QMT exists at very low temperatures around $x$=0.1 with the quantum critical elasticity.

Note that, in the mixed crystals (0$<$$x$$<$1), the chemical disorder induced by mixing must suppress the AFM state and favor the QSL state\cite{31p6,42,43}.
The frustration and disorder effects cannot be separated in this alloying compounds, and it is not possible to determine which effect is dominant.
Nevertheless, both help the present system to show a variation from the FL to the insulating QSL through the genuine Mott transition.
A recent study\cite{43} reveals that disorders do not have a significant influence on the quantum critical scaling characterizing the Mott criticality, and thus, the intrinsic nature of the genuine Mott transition may be discussed in this system even with the chemical disorder.
Also, previous papers on $\kappa$-[(BEDT-TTF)$_{1-x}$(BEDT-STF)$_{x}$]$_2$Cu$_2$(CN)$_3$\cite{37,37p3,37p6,35p5}, in which the ground state changes from the insulating QSL state to the metallic FL state with increasing $x$, also indicate that the disorder effect may not significantly change the Fermi liquidity.
Although the disorder effect must be somewhat related to the electronic states, such as the percolated superconductivity, the present results should reflect the physics near the genuine Mott transition.

\begin{figure}
\begin{center}
\includegraphics[width=\hsize,clip]{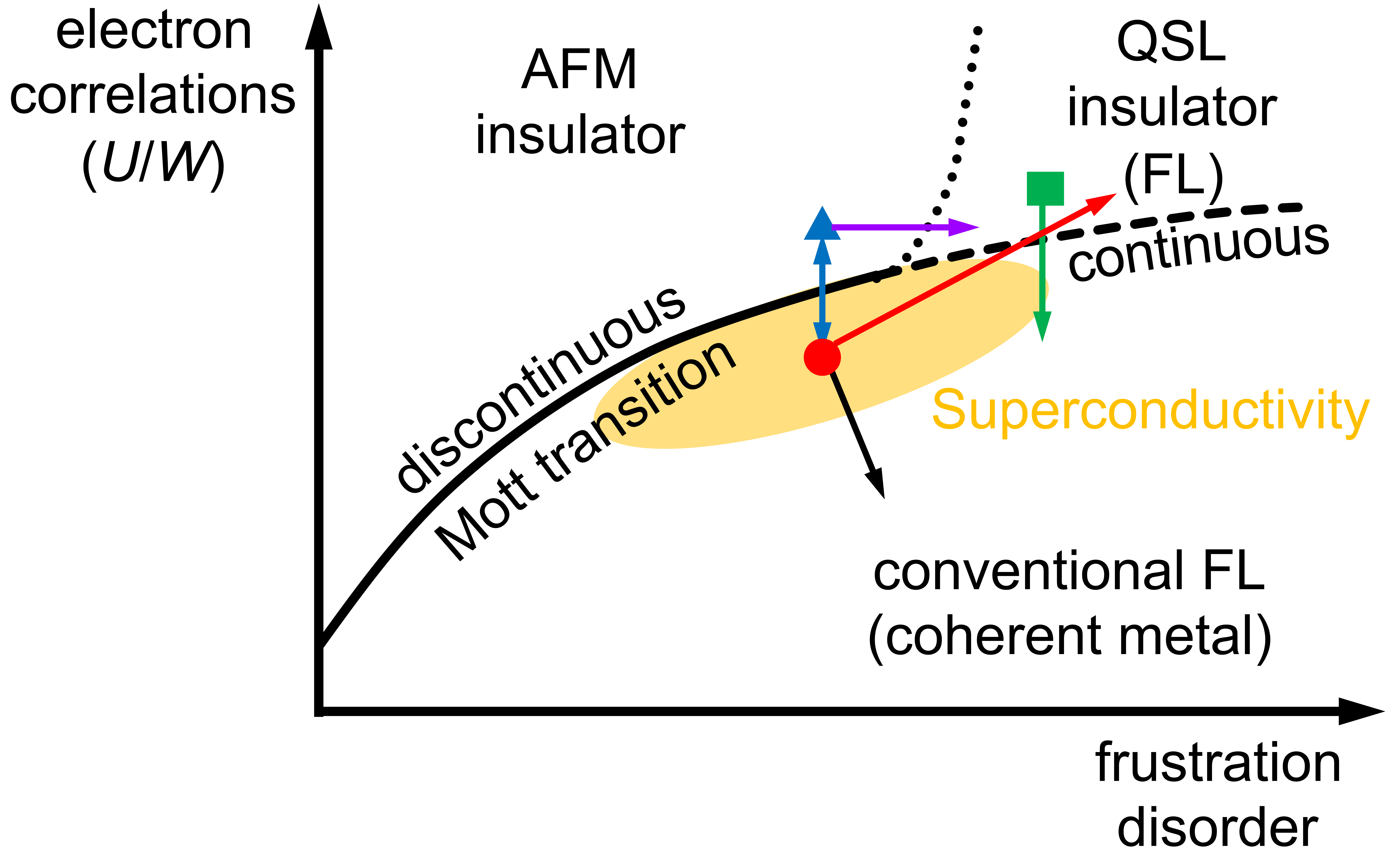}
\end{center}
\caption{
Suggested electronic phase diagram of dimer-Mott organic system.
Each salt is located at symbols (red circle: $\kappa$-(BEDT-TTF)$_2$Cu[N(CN)$_2$]Br; blue triangle: $\kappa$-(BEDT-TTF)$_2$Cu[N(CN)$_2$]Cl; green square: $\kappa$-(BEDT-TTF)$_2$Cu$_2$(CN)$_3$).
Arrows signify routes for changing parameters (red arrow: our study; green arrow: Ref.~\cite{6,20,37}; blue arrow: Ref.~\cite{4,41,22,34,35,36}; violet arrow: Ref.~\cite{42,43}; black arrow: Ref.~\cite{38}).
Orange area indicates superconducting phase in conventional FL region.
}
\label{fig4}
\end{figure}

  We organize our results, the reported experimental data, and the theoretical calculations\cite{39,40} in the form of a generic phase diagram of the ground states in the dimer-Mott system, as illustrated in Fig.~\ref{fig4}.
The present system contains both disorder and frustration as parameters; although they are difficult to separate, they promote the formation of the QSL state\cite{31p3,31p6,42}.
The green box with a downward arrow indicates that the QSL state of $\kappa$-(BEDT-TTF)$_2$Cu$_2$(CN)$_3$ changes into the FL state when external pressures\cite{6,20} are applied or BEDT-TTF is replaced with BEDT-STF (bis(ethylenedithio)diselenadithiafulvalene) \cite{37,37p3,37p6,35p5}.
Our results are indicated by a red circle with an upper right arrow.
The blue arrow represents the numerous studies pertaining to the first-order Mott transition between the AFM and FL states, such as $\kappa$-(BEDT-TTF)$_2$Cu[N(CN)$_2$]Cl under pressure\cite{4,34}, $\kappa$-(BEDT-TTF)$_2$Cu[N(CN)$_2$]Br$_{\rm 1-x}$Cl$_x$\cite{35}, and deuterated $\kappa$-(BEDT-TTF)$_2$Cu[N(CN)$_2$]Br\cite{22,36,41}.
The violet arrow represents recent studies regarding X-ray irradiated $\kappa$-(BEDT-TTF)$_2$Cu[N(CN)$_2$]Cl\cite{42,43}, which shows quantum disordering of the AFM state by randomness.
The black arrow indicates the study for $\kappa$-[(BEDT-TTF)$_{1-x}$(BEDT-STF)$_x$]$_2$Cu[N(CN)$_2$]Br\cite{38}, in which $U$/$W$ decreases as $x$ increases.
The strengthened metallicity reported in Ref.~\cite{38} is consistent with the present description.
The mapping of the universal phase diagram will facilitate the systematic understanding of not only the Mott transition, but also the relation between the magnetic degrees of freedom and the competition among the phases.
Considering the similarity with $\kappa$-[(BEDT-TTF)$_{1-x}$(BEDT-STF)$_{x}$]$_2$Cu$_2$(CN)$_3$\cite{37,37p3,37p6,35p5}, the comparison of the thermodynamic quantities between $\kappa$-[(BEDT-TTF)$_{1-x}$(BEDT-STF)$_{x}$]$_2$Cu$_2$(CN)$_3$ and the present system would be important for a detailed discussion of the effects of disorder and frustration.

  In summary, we investigate the low-temperature states of the dimer-Mott organic system $\kappa$-[(BEDSe-TTF)$_x$(BEDT-TTF)$_{1-x}$]$_2$Cu[N(CN)$_2$]Br.
The genuine Mott transition from the FL state to the QSL state occurs around $x$=0.10 as a possible continuous metal-insulator transition.
We demonstrate that the pure Mott transition results in the freezing of only the charge degrees of freedom with the survival of the spin sector.
The persistence of the fermionic spin excitations is distinct from those of the conventional Mott transition accompanied by AFM orders.

 We thank Dr. A. Kawamoto (Hokkaido University) for advising us about the synthesis of BEDSe-TTF molecule.

\end{document}